\documentclass[sigconf]{acmart}

\AtBeginDocument{%
  \providecommand\BibTeX{{%
    \normalfont B\kern-0.5em{\scshape i\kern-0.25em b}\kern-0.8em\TeX}}}

\setcopyright{acmcopyright}
\copyrightyear{2020}
\acmYear{2020}
\acmDOI{10.1145/1122445.1122456}

\acmConference[Southampton '20]{Southampton '20: 12th ACM Conference on Web Science}{July 07--10, 2020}{Southampton, UK}
\acmBooktitle{Southampton '20: 12th ACM Conference on Web Science,
  July 07--10, 2020, Southampton, UK}

\begin{document}

\title{Dancing to the Partisan Beat: A First Analysis of Political Communication on TikTok}

\author{Juan Carlos Medina Serrano}
\orcid{}
\affiliation{%
 \institution{Technical University of Munich}
 \streetaddress{Richard-Wagner St. 1}
 \postcode{80333}
}
\email{juan.medina@tum.de}

\author{Orestis Papakyriakopoulos}
\affiliation{%
 \institution{Technical University of Munich}
 \streetaddress{Richard-Wagner St. 1}
 \postcode{80333}
}
\email{orestis.p@tum.de}

\author{Simon Hegelich}
\affiliation{%
 \institution{Technical University of Munich}
 \streetaddress{Richard-Wagner St. 1}
 \postcode{80333}
}
\email{simon.hegelich@hfp.tum.de}

\renewcommand{\shortauthors}{Serrano, Papakyriakopoulos and Hegelich}

\begin{abstract}
TikTok is a video-sharing social networking service, whose popularity is increasing rapidly. It was the world's second-most downloaded app in 2019. Although the platform is known for having users posting videos of themselves dancing, lip-syncing, or showcasing other talents, user-videos expressing political views have seen a recent spurt. This study aims to perform a primary evaluation of political communication on TikTok. We collect a set of US partisan Republican and Democratic videos to investigate how users communicated with each other about political issues.  With the help of computer vision, natural language processing, and statistical tools, we illustrate that political communication on TikTok is much more interactive in comparison to other social media platforms, with users combining multiple information channels to spread their messages. We show that political communication takes place in the form of communication trees since users generate branches of responses to existing content. In terms of user demographics, we find that users belonging to both the US parties are young and behave similarly on the platform. However, Republican users generated more political content and their videos received more responses; on the other hand, Democratic users engaged significantly more in cross-partisan discussions.
\end{abstract}




\maketitle
\section{Introduction}

Political communication is contingent on the available information channels. Because social media platforms are a fruitful space for socialization, politics largely takes place on them. Political candidates often exploit such media to interact with the electorate and to place targeted and personalized advertising. At the same time, partisan users utilize their social media accounts  to engage in political discourse. Similarly, a large portion of society obtains its political news updates from social media sources. 

Two factors shape the final forms taken by political communication on social media: \textit{how} a platform  is designed, and \textit{who} uses this space for political purposes. The design of a social media service configures the available information channels for political discourse; its user-base shapes both the generated political content and dictates whether or not a platform can prevail as a significant political space.

Until now, researchers have considered Facebook, Twitter, and YouTube as the most politically relevant social media \cite{tucker2018social}, given the vast number of users who engage daily with these platforms. Nevertheless, social media usage is dynamic. Users change or migrate from one service to another, and some platforms are abandoned as others become popular \cite{kumar2011understanding}. TikTok, a video-sharing social networking service, has recently witnessed a surge in its popularity. It became the world's second most downloaded app in 2019 \cite{sensortower}.

\subsection*{Motivation}

Although researchers have extensively analyzed other popular social media platforms, both by explaining user political behavior and uncovering how platform design influences political communication, no study has focused on TikTok. The present study, therefore, aims to bridge this research gap, intending specifically to understand \textit{who} uses TikTok for political purposes and \textit{how} the design of this platform shapes the flow of political information. For this, we focus on US politics to answer the following research question:

\begin{description}

\item[RQ:]\textbf{What are the features of political communication on TikTok
in terms of (a) partisan users, (b) interaction structure, and (c) diffused content?
}

\end{description}

\subsection*{Original Contributions}

\begin{itemize}

\item We provide a first overview of political communication on TikTok by investigating videos of US Republican and Democratic partisans.

\item We employ computer vision, natural language processing, and statistical tools to evaluate the ways in which partisans combine sound, video, and text to spread their messages. 

\item We study TikTok's interaction design and illustrate how partisans engage in political discussions. We show that TikTok fosters a novel type of political interactivity that is not available on other online social networks. 

\item We investigate user demographics and show that the partisan users are young and behave similarly regardless of their political preferences. We find that Republican users are more active in creating political content and that they receive more responses. We further find that Republicans prefer to engage in video discussions with other Republicans, while Democrats are more open to reach out to users with opposing views.

\item Finally, we discuss other issues related to political communication, privacy, and security on TikTok.

\end{itemize}

\section{Background \& Related Work}

\subsection{Political Communication on Social Media}

The behavior of multiple actors determines the political communication that occurs on social media platforms. Politicians, partisans, and the general public interact constantly, generating complex communication patterns. The design and algorithms of the platforms influence these interactions along with malicious actors to generate a political landscape that can be difficult to understand. Researchers have therefore extensively analyzed various properties of Facebook, Twitter, YouTube, and other similar social media spaces to grasp the nature of politics on social media.

Several studies have investigated how politicians use social media for political purposes. They have shown that politicians use platforms in different ways, depending on the audiences and sociotechnical environments \cite{stier2018election, enli2013personalized, stieglitz2013social, serrano2018social}. They have also explained how services are used for personalized advertising campaigns \cite{papakyriakopoulos2018social} and whether the presence of politicians on these platforms affects their electoral popularity \cite{effing2011social}.

Other studies have explicitly focused on user political behavior. They have investigated how partisans of different political orientations use social media platforms \cite{Bode2016, serrano2019rise}, analyzed how user activities are generally distributed \cite{papakyriakopoulos2020political} and evaluated the spread messages' content and polarity \cite{stieglitz2012political, ottoni2018analyzing, chatzakou2017mean, papakyriakopoulos2020bias}. Studies have also focused on the usage of social media in periods of social unrest \cite{varol2014evolution}, analyzed online social platforms as spaces for the coordination of social movements \cite{tufekci2012social}, investigated how different social groups behave, and scrutinized the conditions in which they help to polarize and segregate the citizenry \cite{bail2018exposure, barbera2014social, conover2011}. 

Inseparable to political communication on social media is news consumption, with a large proportion of the public using the social media services as their primary information source pertaining to world events. Many studies have investigated agenda settings effects, as well as how different types of news coverage affect user behavior and contribute to attitude formation \cite{flaxman2016filter, boczkowski2018news, king2017news}. Moreover, researchers have also extensively studied how low credibility news and misinformation are diffused on social media platforms by real, fake, or automated accounts \cite{vosoughi2018spread, howard2016political, ferrara2016rise, hegelich2016social, del2016spreading, allcott2019trends}. 

An equally important role in understanding user behavior is the analysis of the content filtering algorithms employed by social media services. Thus, researchers have investigated social platforms as algorithmic ecosystems and have studied whether and how recommendation algorithms influence the public's political behavior and opinions \cite{ribeiro2020auditing, papakyriakopoulos2020political, bakshy2015exposure, lustig2016algorithmic, niemanlab}. In the same vein, scholars have evaluated platform design features and how they shape information diffusion \cite{stier2018election, enli2013personalized,malik2016identifying}. These dimensions of political communication have been extensively analyzed for the US on different social media platforms \cite{allcott2017social, grinberg2019fake, kreiss2018technology, wells2016trump, hall2018brexit, enli2017twitter}, but not yet on TikTok.

\subsection{TikTok}
TikTok was created by ByteDance, a Beijing-based tech company. The company had previously launched Douyin for the Chinese market in September 2016. Subsequently, the company launched TikTok in 2017 for markets outside China. Both services are similar, but they run on separate servers to comply with China's regulations. In 2018, TikTok merged with the social media app Musical.ly to create a larger video community.  In October 2019, TikTok and Douyin jointly achieved 800 million monthly active users \cite{businessinsider}. In the United States, 60\% of TikTok's 26.5 million monthly active users are aged between 16 and 24 \cite{reuters}. The platform is mainly accessible through a mobile app. Although it is possible to access posted videos from non-mobile devices, the features are limited as it is not possible to create content or read user comments.

TikTok offers users a unique method of sharing creative videos of themselves, their surroundings, or a compilation of external audiovisual content. The simplest videos consist only of text superimposed onto a colored background.  Videos can then be more complex by including images, video clips, and sounds. The images and video footage can be altered using the app's voice effects, image filters, and video speed controllers. The maximum length of a video post is 60 seconds, and they can consist of a collection of shorter video clips that tell a story when they are combined. When the users post videos, they can add a caption with hashtags to describe their clips. Like Twitter, the most used hashtags represent topics that are trending on the platform, and like Instagram, the video clips are classified according to their hashtags. 

TikTok is considered a social media platform because like Twitter and Instagram, its users have a social group of followers and other users they follow. However, the main feature that differentiates TikTok from other social media services, is the videos' background music, which represent the core message that the users want to convey.  Users can select background music for their videos from a wide variety of music genres and can even create original sound clips. Any sound clip, including user voice messages, can be selected by other users to use in their videos. For many videos, the music serves as part of a dance routine, a lip-synching battle, or as the backdrop for a comedy skit. However, sound can also function as a story builder and can be used to deliver a specific message. For example, a famous original sound clip begins with La Roux's song, Bulletproof, remixed by Gamper \& Dadoni. Then the music stops, four gunshots are heard, and a man's voice says "there's not any". Text messages appear in the part of the clip that plays the song, citing reasons why a particular cause should be supported, and the user points at the messages. When the gunshots begin, however, the user makes a gun sign with a hand and shoots at the text snippets containing the reasons for support. The user is implying that the citing reasons are invalid and that there is no real reason for people to support that cause. 

Users consume content by viewing an algorithmically generated feed of videos on the so-called "For you" page, which is the landing place when users open the app. Although there is no explanation of how the algorithm work, the videos that appear to the user largely rely on a central recommendation algorithm instead of on the activities of the user's social network \cite{buzzfeed}. According to TikTok, the "For you" page is "a personalized video feed specifically for you based on what you watch, like, and share."\footnote{https://apps.apple.com/us/app/id835599320} This contrasts with Facebook's and Twitter's feed, which rely mostly on the user's social graph and resemble more YouTube's recommendation system. Users can also search for hashtags, sounds and find the trending videos on the "Discover" page. 

\begin{figure}[htp]
\centering
\includegraphics[scale=0.2]{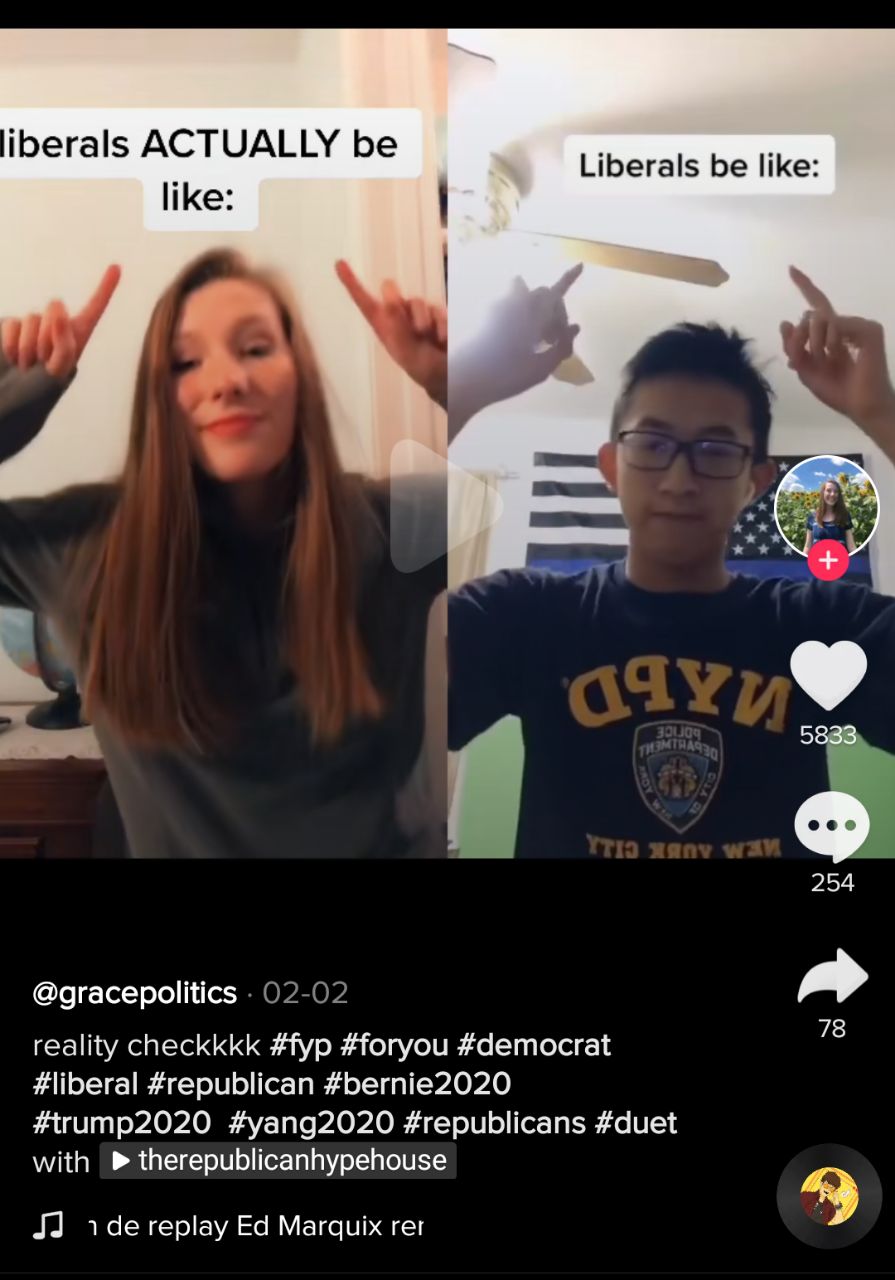}
\caption{Screenshot of a duet on TikTok. On the right is the original video and on the left, the video posted in response.}
\label{duet}
\end{figure}

A unique feature on TikTok is the duet. A duet is a response video that allows users to reply directly to a video post with a video of their own. The original and duet videos play side by side and the music clip from the original video's audio is preserved. Since the audio does not change, the duet exhibits users responding through text snippets, images, or facial expressions. Users can also create a reply duet from an existing duet. In such instances, three videos appear together.  Figure \ref{duet} shows a screenshot of a duet on TikTok. The original video is placed on the right and the duet video on the left.  The screenshot displays the number of likes, comments, and shares the duet video attracts. The music that appears on the bottom is a music clip from a remixed song. In the image, both users point to a text snippet and communicate their perspectives as they dance to the same music. In this way, TikTok can be used to share opinions on controversial topics. In this study, we focus on political content to determine how TikTok users interact and show their support for a political party with the help of music, video, and image. 

\section{Data \& methods}
\subsection{Data Collection}

TikTok allows users to search for videos with a specific hashtag and view the most popular results. We decided to crawl the videos containing the hashtags \textit{\#republican} and \textit{\#democrat} on February 1, 2020. The hashtag search yields a limited number of videos and it is not clear how this limit is defined. Popularity may play a role as a hashtag search showcases the most popular videos. For the two hashtag queries, we obtained a different number of videos. The collection process resulted in 3,310 videos: 2,362 with the hashtag \textit{republican}, and 1,831 with the hashtag \textit{democrat}. Of the total, 350 were duets; thus, we also collected the original videos from them if they were not yet in the data. To expand the dataset, we searched for duets to the videos we had collected. Unfortunately, TikTok does not offer a search by video feature that directly links a video to its duets. However, it is possible to search by sound clip. This search shows videos that have employed the same sound. Our approach was to search for the sound of each video and to add the videos that were dueting to the videos in our dataset. As with search by hashtag, only a limited number of results can be obtained. This presents a limitation to collecting all the duets to a video, especially for videos that use extremely popular sounds. Nonetheless, searching for original sounds often yielded only the duets of the given video. After this procedure was complete, our dataset consisted of 7,825 TikTok videos. Most of the videos were created between October 2019 and February 2020. The oldest included video was posted in March 2019. 

Before beginning the analysis, we manually labeled every original video and duet as pro-Republican, pro-Democrat, or nonpartisan. This coding was conducted by two of the main authors. Both authors labeled each video individually. For the cases where authors disagreed, the third author was consulted to resolve the coding conflict. Videos that directly supported or opposed a political party or a member of a party were classified accordingly. We labeled videos opposing one Democratic candidate but expressing support for another candidate of the same party as pro-Democrat. Videos in which users articulated their standpoints on issues such as abortion, guns, and LGBT rights, were not directly attributed to any political party. We only classified those videos that indicated a clear political affinity toward the Republican or Democratic parties as partisan. For example, social issue videos with both the \textit{republican} and \textit{democrat} hashtags were coded as nonpartisan. In total, we classified 5,946 videos as partisan posts. Apart from assigning partisanship, we grouped the partisan videos into four categories: 1) videos that included the user's face; 2) videos where the user appears but does not show its face or videos culled from other sources such as news clips; 3) videos solely comprised of images; and 4) videos that only showcased textual content. We used the same coding procedure as the classification of partisan videos. It was important for us to view all the videos to be able to understand TikTok's communication structures and the user behavior displayed on the platform. 

\subsection{Methods}
TikTok videos are rich in features and extra pre-processing steps were required to extract the information for analysis. From the original videos, we started by extracting the text snippets from the videos. For this, we divided the videos into images every two seconds and then employed Tesseract, an optical character recognition engine. From the original videos that included the user's face, we used the images and processed them via Microsoft's Azure Face API\footnote{https://azure.microsoft.com/en-us/services/cognitive-services/face/}, which allows emotions, gender, and age to be extracted. Additionally, we employed IBM's text to speech API\footnote{https://www.ibm.com/cloud/watson-text-to-speech} to extract the audio from the videos that contained original sounds created by the users. 

To explore the difference in the usage of hashtags, we employed a measure of political valence introduced by Conover et al. \cite{conover2011}. 
They defined political valence $V$ of a hashtag $h$ as
$$V(h) = 2 * \frac{\frac{N(h, R)}{N(R)}}{\frac{N(h, D)}{N(D)} + \frac{N(h, R)}{N(R)}} - 1 \quad \quad (1)$$
where $N(h, D)$ and $N(h, R)$ indicate the number of appearances of a hashtag, and N(D) and N(R) represent the total number of hashtags in the Democratic and Republican videos, respectively. This equation bounds the valence between -1 for hashtags only used in Democratic videos and +1 for hashtags appearing only in Republican videos.

We created a graph of interactions where the users are the nodes and the directed edges represent duet interactions. This graph allows to measure the extent of cross-ideological (R\textrightarrow D, D\textrightarrow R) and intra-party interactions (R\textrightarrow R, D\textrightarrow D). We used another measure from Conover et al., which divides the observed number of interactions with the expected number of interactions. The expected value assumes that the source of the edge is preserved and the target of the edge is randomly assigned to one user, irrespective of the individual's political orientation. For example, the expected interactions between Republicans and Democrats is defined as:
$$E[R \longrightarrow D] = k_R * \frac{Users_D}{Users_D + Users_R} \quad \quad (2)$$
where $k_R$ refers to the number of edges originating from pro-Republican videos and $Users_D$ denotes the number of Democratic users. 

Finally, we applied topic modeling to evaluate the content of video captions. We used a Latent Dirichlet Allocation algorithm \cite{blei2003latent} to extract the latent thematic topics and to calculate the empirical distribution of videos belonging to the identified topics $f(video|topic)$. Using this distribution, we computed and compared the amount of Democratic and Republican videos associated with each topic. In order to calibrate the number of topics and the latter model hyperparameters, we perform sensitivity analysis for various values and select the model with the highest topic coherence score, as suggested by R\"oder et al. \cite{roder2015exploring}.

\section{Results}
We illustrate different features of political communication on TikTok (\textbf{RQ}) in the following three subsections. First, we describe the general statistics of the collected dataset, including user demographics and activities. Second, we describe and evaluate political interactions between partisan users. Finally, we analyze the content of different information channels used in TikTok videos to detail the nature of the political discourse.

\subsection{Descriptive Statistics}
Of the 5,946 partisan videos, 2,802 are original videos and 3,144 are duets. Table \ref{statistics} shows the total number of videos classified as pro-Democrat or pro-Republican and the extent of the interactions generated by these videos. In our dataset, there are two times more Republican videos than Democratic videos. Overall, the Republican videos accumulate more likes, shares, and comments. We applied one-sided Mann-Whitney tests for each reaction to compare if there is a significant difference between the partisan videos. For likes, the Republican median is 497 and the Democratic median is 232 (p < 0.00). In terms of shares, the Republican median is 6 and the Democratic median is 3  (p < 0.00). With regard to the comments, the Republican median is 19 and the Democratic median is 13 (p < 0.01). All the tests are significant and show that Republican videos attracted more interactions in general. 
\newcommand{\ra}[1]{\renewcommand{\arraystretch}{#1}}

\begin{table}[!h]\centering
\ra{1.3}
\caption{Number of videos created by pro-Republican and pro-Democratic users and user engagement (likes, shares, and comments) with them.}
\begin{tabular}{@{}lrrrrr@{}}\toprule

  & \textbf{Videos} & \textbf{Users} & \textbf{Likes} & \textbf{Shares} & \textbf{Comm.}\\ \midrule
\textit{Republican} &3,987&1,957&15,533,963&817,728&500,514\\
\textit{Democrat} &1,959&1,249&10,663,139&392,468&257,199\\
\bottomrule
\end{tabular}

\label{statistics}
\end{table}

\begin{figure}[htp]
\centering
\includegraphics[width=8.2cm]{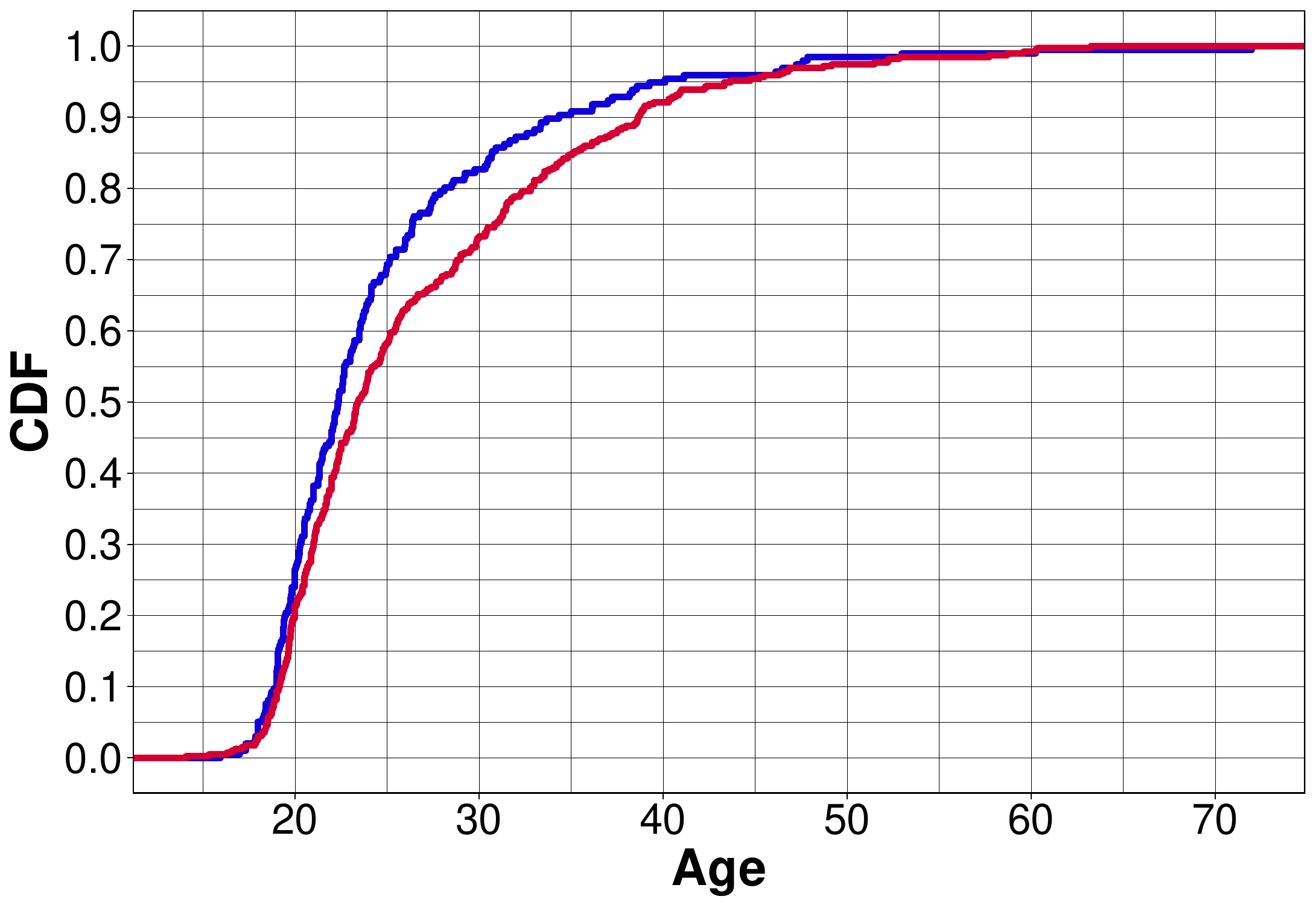}
\caption{Cumulative distribution of the users' age divided as Democratic (blue) and Republican (red) users.}
\label{cdf}
\end{figure}

Of the original videos, 70\% included the user's face, 22\% featured other video content, 7\% comprised only image content, and 1\% only exhibited text. We performed feature extraction for the original videos that included the user's face. To this end, we divided the videos into pictures and obtained the features for each picture. With Microsoft's API, we were able to extract gender, age and emotions, which include anger, contempt, fear, happiness, sadness, and surprise. Afterward, we averaged the emotions and ages of all the pictures to obtain the mean values for each video. We used the mode for gender given that it is a categorical feature. For age and gender, we aggregated the videos per user to obtain unique values. We manually categorized the users, for which the mode of the gender was inconclusive. From the original Republican videos, 219 users are male and 187 are female, whereas for the Democratic videos, 84 users are male and 118 are female. We perform goodness of fit chi-squared tests to evaluate the male-female user balance. For the overall population, gender is balanced ($\chi^2$=0.01, p=0.935). The same applies for the Republican partisans ($\chi^2$=2.52, p=0.11). However, a significantly larger number of Democratic partisans posting original videos are female ($\chi^2$=5.72, p=0.016). Figure \ref{cdf} portrays the cumulative age distribution of Democratic and Republican users. We observe that in general, the Democratic users are younger than Republican supporters. The percentage of Republicans between 16 and 24 years old is close to 60\%, which mirrors the percentage of US TikTok users in the same demographic group. For the Democratic users, however, this percentage is closer to 70\%, younger than the mean user age. Nevertheless, the majority of users creating political content are below 40 for both parties in our data sample.

Table \ref{emotions} illustrates the average emotion expressed for the posted videos per party. We do not observe significant differences between the two groups. We conclude that users on TikTok express themselves similarly irrespective of the party they support. Interestingly, happiness and surprise have higher averages than anger or sadness. We presume that this finding is related to the nature of TikTok comedy skits. Indeed, we observed many videos that relied heavily on sarcasm: smiling and dancing individuals confronting users supporting the views of the opposite political party and mocking or belittling them.

\begin{table}[]\centering
\ra{1.3}
\caption{Average emotion expressed by TikTok users divided by partisanship.}
\begin{tabular}{@{}lrr@{}}\toprule
\textbf{Emotion}         & \textbf{Democrat} & \textbf{Republican} \\ \midrule
anger     & 0.021    & 0.019      \\
contempt  & 0.013    & 0.014      \\
fear      & 0.004    & 0.004      \\
happiness & 0.217    & 0.212      \\
neutral   & 0.619    & 0.635      \\
sadness   & 0.047    & 0.044      \\
surprise  & 0.074    & 0.067     \\
\bottomrule

\end{tabular}

\label{emotions}
\end{table}

\subsection{Interaction Structure} 
Political communication on social media is influenced directly by the design of the platform by determining the interaction structure between users. The interactions can be ordered in consecutive levels of communication, with each increasing level representing a more direct response. We identify four levels of communication on TikTok. The first level corresponds to an indirect response when a user views a video. Although there is no active reaction from the user, the message is received and processed. Additionally, from the data perspective, the view counter on the video increases, and this metric can influence TikTok's recommendation algorithm.  The second level of communication consists of a basic response that involves liking the video or sharing it. The next level constitutes written responses to a video through user comments.  On other social media platforms, this is the highest level of user response to a posted element. On TikTok, however, a fourth level of communication allows users to respond with a video, a feature referred to as duet. The duet shows more similarities to face-to-face communication than to a written response, which can make the interaction between users feel more personal.   

\begin{figure}[h!]
\centering
\includegraphics[width=8cm]{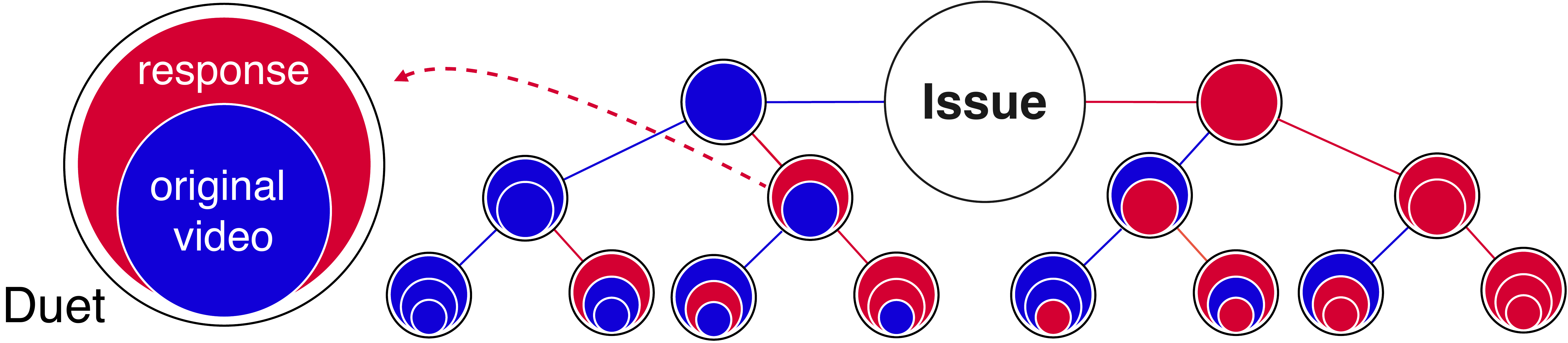}
\caption{Communication tree for TikTok duets.}
\label{tree}
\end{figure}

The structure of duets directly affects how political communication takes place on TikTok. The duets follow a tree structure, where users create branches by responding to other videos. We depict this tree structure of communication in Figure \ref{tree}. On top of the tree, there is a political issue, which partisan users use as their motive for the production of pro-Democrat or pro-Republican videos. Connected on the second depth level are the original content videos. The third depth level represents the duets  to the original videos. The next level nodes on the tree denote duets posted in response to previous duets. It is possible to continue to react with duets further than the three duet depth levels displayed in Figure \ref{tree}. Inner nodes on the duet nodes have been included in the illustration to represent that duets, previous duets, and original videos appear side by side on TikTok. A user interacting with a deep level duet video can directly observe the complete communication chain on-screen without needing to scroll down. This differentiates duets from any feature available on other social media networks.

Using the manually labeled videos, we were able to quantify the duet interactions between partisan users. Table \ref{interactions} shows the percentage of partisan and cross-ideological interactions. It also includes the observed interactions divided by the expected interactions as presented in Formula 2. We observe that 77\% of the Republican duets represent responses to Republican users. In contrast, more than 80\% of the Democratic duets were directed toward Republican supporters. This inverted dynamic also appears in the ratio between observed and expected interactions. Democrat-Republican and Republican-Republican interactions present a ratio larger than one (1.35 and 1.28 respectively). We include the ratio, as it is a standardized measure that can be used to compare to the outcomes of other studies. For example,  Conover et al. \cite{conover2011} evaluated political communication in the US on Twitter and found that retweets had higher ratios than one for intra-partisan interactions (1.70, 2.32) and lower ratios than one for cross-partisan interactions (0.03, 0.05). Mentions displayed a similar but less pronounced effect for both parties (1.23, 1.31 for partisan exchanges, and 0.68, 0.77 for cross-ideological interactions). Therefore, the authors found that user behavior with regard to Twitter mentions and retweets was unrelated to the political party. In contrast, we find that TikTok duets represent a party-specific structure. This difference shows the importance of studying the effects of different platform designs on the political communication that occurs between users.

\begin{table}[!h]\centering
\ra{1.3}
\caption{Interactions between partisan and cross-ideological users, including the percentage and ratio between observed and expected interactions.}
\begin{tabular}{@{}lrrrr@{}}\toprule
&\multicolumn{2}{c}{\textbf{Percentage}}&\multicolumn{2}{c}{\textbf{Ratio}}\\
& \textbf{\textrightarrow D} & \textbf{\textrightarrow R} & \textbf{\textrightarrow D} & \textbf{\textrightarrow R} \\ \midrule
Democrat     & 19\%    & 81\%   &0.48 &1.35   \\
Republican  & 22.6\%   & 77.4\% & 0.57 & 1.28    \\
\bottomrule

\end{tabular}

\label{interactions}
\end{table}

\begin{figure}[h!]
\centering
\includegraphics[width=8cm]{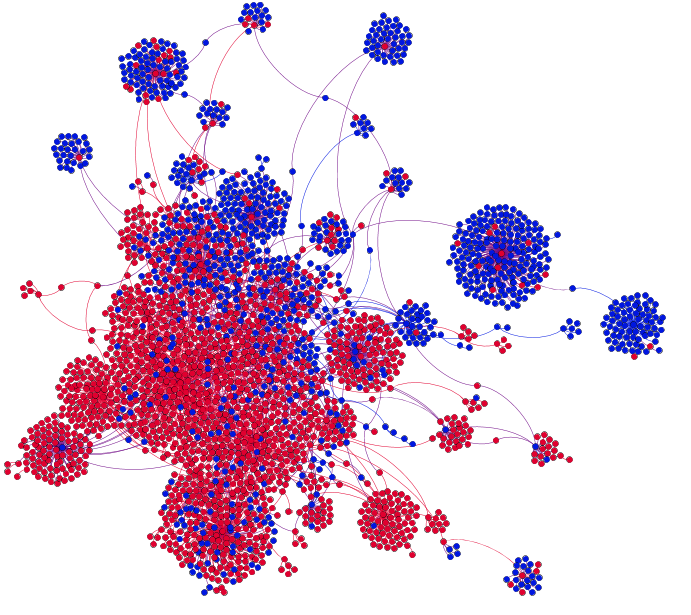}
\caption{Graph of duet interactions between partisan users. Red nodes correspond to Republicans and blue nodes to Democrats. Purple edges depict cross-ideological exchanges.}
\label{network}
\end{figure}

We portray the duet interaction between users in Figure \ref{network}. Each node represents a user and the edges represent two users of a duet. Blue nodes indicate Democrats and red nodes designate Republicans. We only find one account that posted both pro-Democrat and pro-Republican videos, and this account was omitted from the analysis. The graph shows a tight Republican cluster in the middle with some Democratic users interacting with this community. The boundaries evince large clusters of Democratic users responding to Republican accounts. These users are separated from the main cluster because they did not interact with accounts other than a specific Republican user. The graph confirms the result of high partisan interactions among Republicans and the high cross-ideological interactions from Democrats to Republicans.

\subsection{Content Analysis}

The content analysis of the various information channels on TikTok provided important insights on how political communication takes place on the platform. Each information channel was used differently by partisans. Partisan users generally avoided political statements in their profile descriptions, except for some users whose username explicitly stated their political affiliations. Most users added links to their social media handles on other platforms such as Twitter or Instagram,  and some provided their Venmo accounts for fans to support them financially.

\begin{figure*}[!h]
\centering
\includegraphics[width=17.2cm]{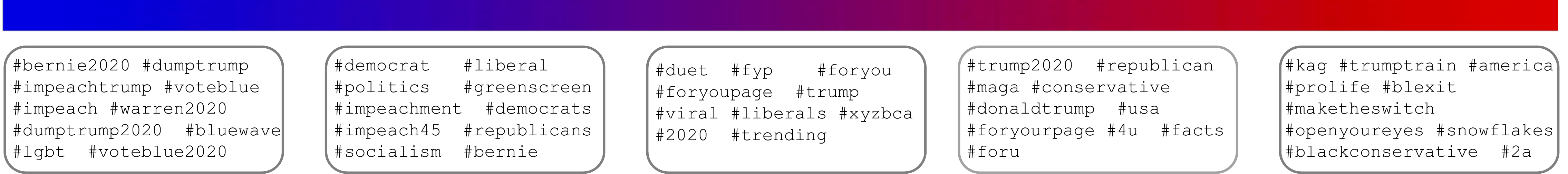}
\caption{Political valence of hashtags on the Democratic-Republican spectrum.}
\label{valency}
\end{figure*}

In contrast to the profile descriptions, video captions were extremely politicized. Users usually inserted as many political hashtags as possible from across the political spectrum, to ensure the visibility of their content. Their hashtag selections were also influenced by partisanship. Users inserted specific hashtags more often in accordance with their political orientation. To reveal this, we divided the valency spectrum in five equidistant groups and grouped the hashtags according to valences.   Figure \ref{valency} presents the top ten words per group.  Democratic partisans more frequently used hashtags related to the impeachment of Donald Trump, Bernie Sanders, Elizabeth Warren, and LGBT issues. Republican partisans more often used hashtags related to Trump's campaign slogans and phrases used by the alt-right to assign credibility to information such as \#facts, \#maketheswitch, and \#openyoureyes. Regardless of partisanship, users added to their videos platform-specific hashtags such as \#foryou, \#foryourpage, and \#xyzbca. These hashtags are irrelevant to the political discussions but constitute a cardinal aspect of TikTok interactions.

\begin{figure}[b!]
\centering
\includegraphics[width=8cm]{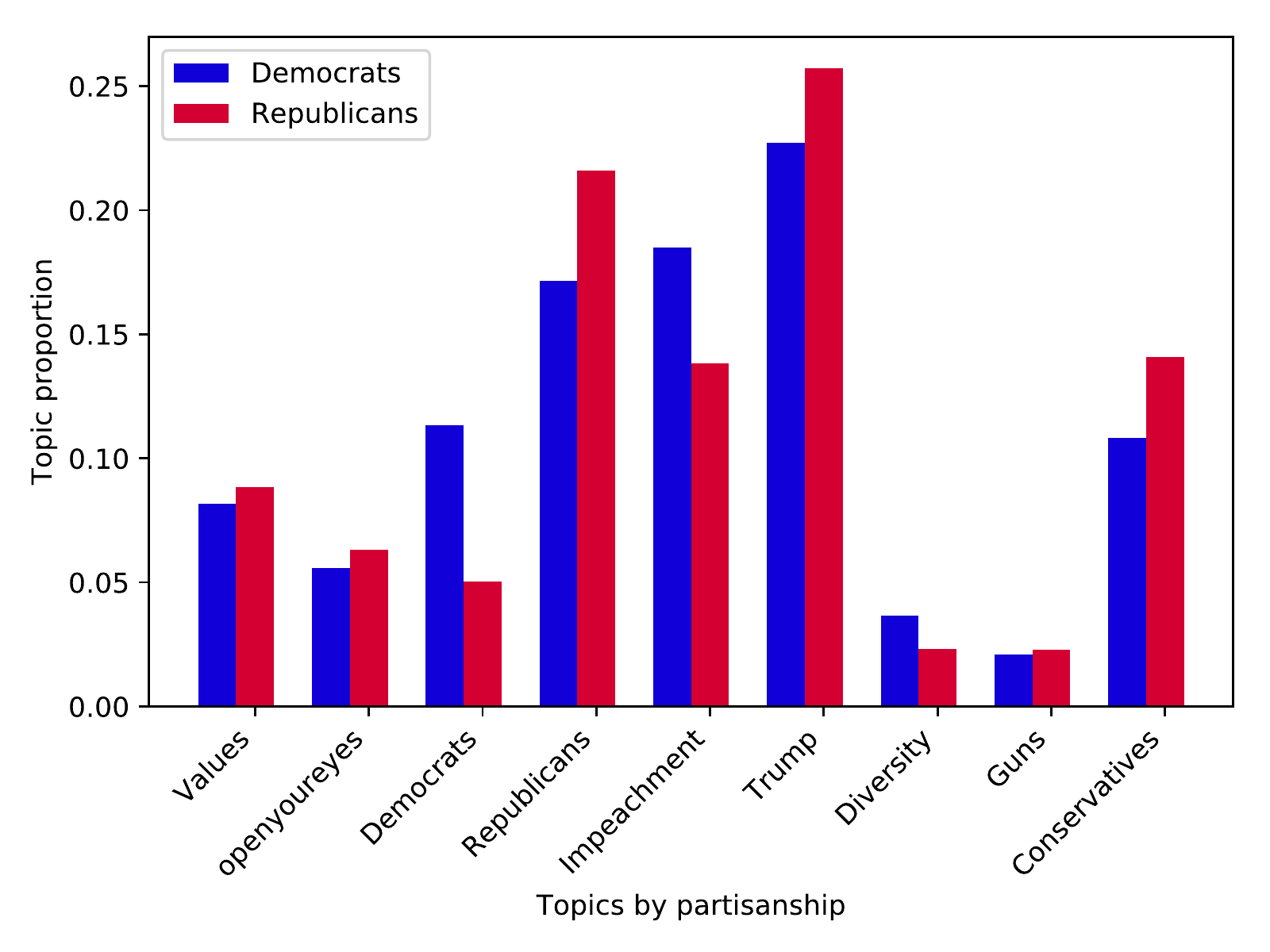}
\caption{Ten predominant topics appearing in videos captions on TikTok and their distribution between Democratic and Republican users.}
\label{topics}
\end{figure}

\begin{figure}[htp]
\centering
\includegraphics[width=8.2cm]{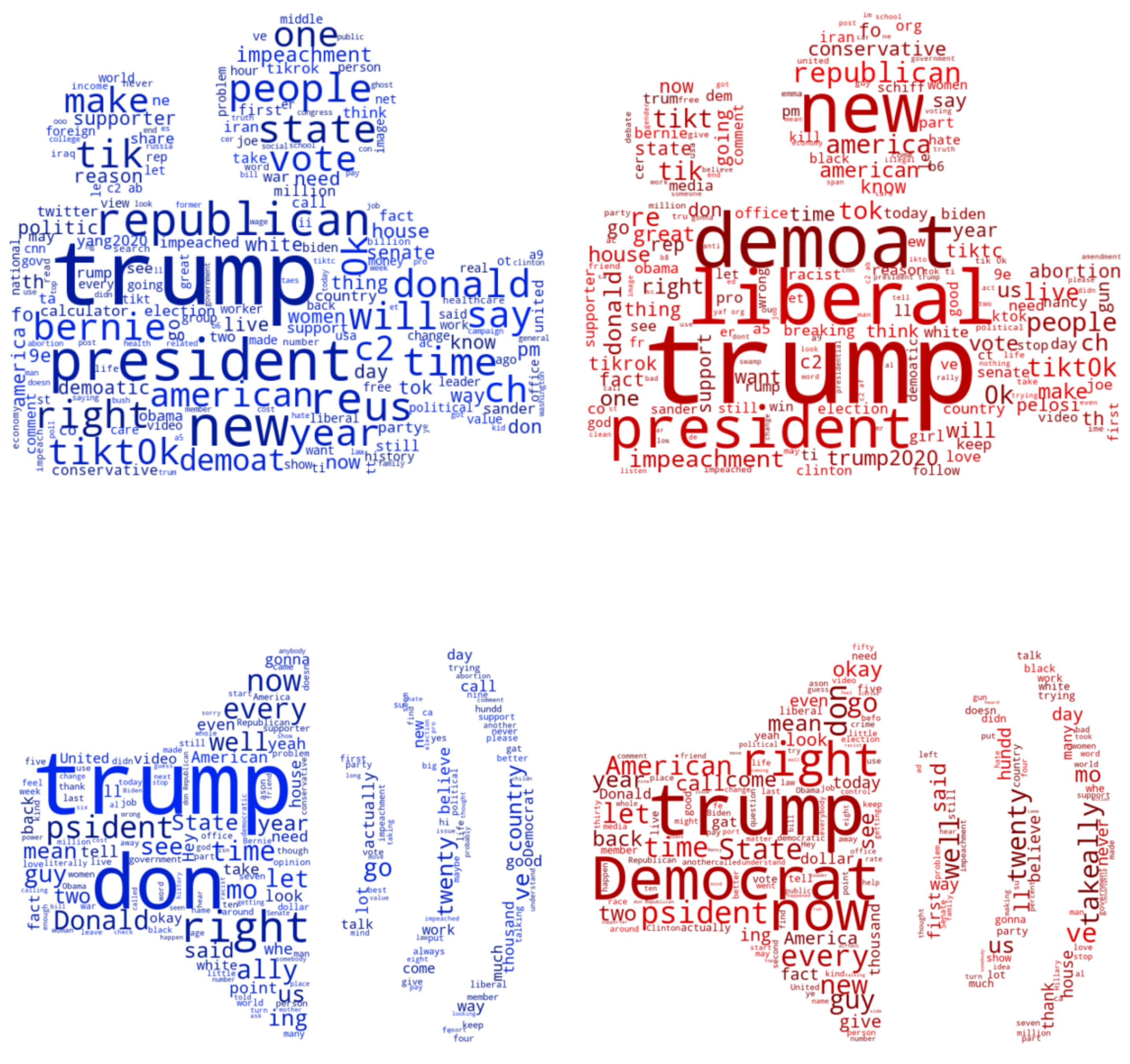}
\caption{Most frequent words appearing in embedded texts snippets (up) and user audio messages (down).}
\label{wordcloud}
\end{figure}

The topic modeling algorithm provided more detailed information about partisan interests for the specific period on TikTok. The model optimization process for the video captions yielded ten topics on which both Republican and Democrats generated content. Figure \ref{topics} shows that although some topics were more prevalent in posts by Democrats or by Republicans, both groups engaged in the same discussions. The topics that were discussed more or less equally by both sides related to social issues such as religion and abortion, guns and the second amendment, as well as discussions associated with daily political developments. Democrats generated content related to their party, about Trump's impeachment trial, as well as social diversity. In comparison, Republicans created more videos about their party, conservative values in general, and about Donald Trump's activities.

The analysis of audio and the text embedded in the videos evidenced particular ways in which these communication channels were used (Figure \ref{wordcloud}). Both Republican and Democratic partisans used sound and embedded text to call out their opponents, setting the stage for the audiovisual discourses. Donald Trump was the focal character of these channels and was most frequently mentioned in the videos. Nevertheless, the embedded text contained additional information about the opinions of the partisans, and often mentioned other political candidates, expressing support for them or criticizing them. Furthermore, the text also included issues of interest to the partisan users, such as Donald Trump's impeachment, abortion rights, healthcare, and the second amendment. These results illustrate that partisans used audio, video, captions, and user descriptions in different ways, creating a complex multichannel information flow in their political interactions.

\begin{table}[!b]\centering
\ra{1.3}
\caption{Number of total views for TikTok videos with a given hashtag. Number of Instagram posts that include the same hashtags.}
\begin{tabular}{@{}lrr@{}}\toprule

\textbf{Hashtag}  & \textbf{TikTok Views} & \textbf{Instagram Posts}\\ \midrule
\textit{\#foryou} &1,687B&-\\
\textit{\#foryoupage} &968B&-\\
\textit{\#trump} &730.3M&13M\\
\textit{\#trump2020} &1.1B&1.2M\\
\textit{\#bernie} &34.8M&564K\\
\textit{\#bernie2020} &166.3M&216K\\
\textit{\#biden} &4.8M&102K\\
\textit{\#biden2020} &1.9M&26.9K\\
\textit{\#warren} &3.5M&253K\\
\textit{\#warren2020} &11.6M&38.9K\\
\textit{\#billieeilish} &3.5B&5.6M\\
\textit{\#shawnmendes} &1.4B&9.5M\\
\textit{\#gretathunberg} &100.5M&381K\\
\bottomrule
\end{tabular}

\label{instagram}
\end{table}

Our final analysis assessed the extent of political content on TikTok in comparison to non-political content. TikTok's search tool reports the number of total user views for videos that include a particular hashtag. This allowed us to compare between political and non-political hashtags. Thus, we searched for a selected number of hashtags on February 1, 2020. Table \ref{instagram} shows hashtags of several political actors, including their names and their names plus \textit{2020}. We also include three popular personalities and the most popular hashtags on TikTok, \textit{\#foryou} and \textit{\#foryoupage}, for comparison. Among the political hashtags, \textit{\#Trump2020} leads by a substantial margin, with a total of 1.1 billion views. The political hashtag that comes second with 166.3 million views is \textit{\#Bernie2020}. Interestingly, the hashtags with the names of the candidate and \textit{2020} have more views than hashtags that included only the candidate's name. The large difference between \textit{\#Trump2020} and the rest of the Democratic candidates corroborates our finding pertaining to the greater number of pro-Republican videos seen in our data and evidences the absence of a possible sample bias effect in the collection procedure. Apart from politics, hashtags referring to singers Billie Eilish and Shawn Mendes have total views comparable to Trump-related videos, whereas Greta Thunberg videos have the number of views in the same order of magnitude as the videos related the Democratic candidates. The most famous hashtag on TikTok has 1,687 more views than \textit{\#Trump2020}. As a rough cross-platform comparison, we include the number of Instagram posts for the same hashtags in Table \ref{instagram}. Instagram's search tool displays the number of posts for a hashtag but not the number of views. The number of posts cannot be compared to the number of views as they represent two different quantities. The proportions on Instagram between politicians and popular personalities are similar to the proportion in views on TikTok. However, there are more videos with only the politicians' names than hashtags including \textit{2020}. This result could signify that there is more content focused on the 2020 US presidential campaign on TikTok than on Instagram. In sum, we conclude that not only does US political content takes place on TikTok, but it also accounts for a large ecosystem on the platform.

\section{Discussion}

The results of this study demonstrate that a new form of political communication takes place on TikTok. Communication still preserves its decentralized character as on  other social media platforms, with users generating, sharing, and diffusing information. However, TikTok users do not just merely circulate content and comment it; they \textit{become} the content. In contrast to Facebook and Twitter, where users exchange news articles in the form of URLs and articulate their political opinions through comments or feedback posts, TikTok users become active presenters of political information. Every TikTok user is a performer who externalizes personal political opinion via an audiovisual act, with political communication becoming a far more interactive experience than on YouTube or Instagram. Since every user seeks increased popularity to disseminate their messages more widely, they create short political spectacles resulting in the realization of \textit{politics as entertainment}. Unlike television media where news anchors and political pundits are the showmen and women, everybody on TikTok is one. It is not a coincidence, therefore, that this intensive audiovisual universe attracts young users who actually "play" their politics on the platform.

The duet function is one of the main reasons why political communication is so interactive on TikTok. Users can employ a variety of elements to respond to  videos posted by other members. These features range from simple facial reactions to text snippets that serve as fact-checking points. Some users even overwrite the original video's text to "correct" the other user's stance on a topic and showcase opposing arguments. Moreover, the audience interacting with duet videos can directly compare the different points of view. This duet structure contrasts with other social media platforms where public exchanges take the form of written responses that appear as a list under the original post. Duets also allow users to exhibit their creativity, in showing support or making counterarguments. For these reasons, we argue that duets are the closest feature on social media to an actual online public debate.

Given that TikTok's design introduces a novel way of conducting politics, it is reasonable to ask how this framework can transform other aspects of political communication. In this study, we illustrated how political partisans generate content and interact on TikTok. However, multiple other dimensions of political communication should also be investigated. For example, although news media URLs are not diffused as in other social platforms, many news media agencies already have TikTok profiles to broadcast reports to the public. The same applies to a handful of political candidates who use TikTok as a new medium of reaching the electorate. Recently, TikTok followed Twitter's suit and banned the placement of political advertisements on the platform \cite{newsroom}. These phenomena and decisions interfere directly with political campaigning and opinion formation; therefore, researchers should investigate these actions more comprehensively in the future. This also applies to researchers who plan to study the general user behavior on the platform. Scholars should seek to uncover whether the platform design and the deployed recommendation algorithms result in the polarization or segregation of social groups, and whether hyperactive user behavior has an agenda-setting effect on the platform. Although TikTok primarily involves real users who reveal themselves in front of the camera, it is equally important to study whether and how any misinformation attempts take place, as well as how users present controversial issues on the platform.  Researchers should undertake the task of determining whether offensive and discriminatory content prevails on the platform.

Besides the aspect of political communication, further political issues regarding user privacy and security should be addressed. Given TikTok's open nature, such concerns for the users have already been raised \cite{thehill} but require more in-depth evaluation. Although users can create private videos visible only to their friends, the platform is mainly geared toward the production of viral videos. This means that data is easily reachable for data mining processes. Indeed, TikTok is a rich information source because its content reveals the personal features of users through the immediacy of audiovisual media to their appearances, personalities, traits, vocal attributes, and points of views. Moreover, users creating and interacting with political content can be classified by their partisanship as we did in this study. With the manually classified videos, a machine learning algorithm may be trained to identify political content and to automatically assign partisanship to a TikTok user. This information can then be employed for political or advertising purposes as it is already prevalent on other platforms like Facebook \cite{nytimes}. However, the potential risks are higher on TikTok because advances in facial recognition technologies make it possible to identify individual users and match them with citizenship records. Although this scenario is also possible on other platforms, active TikTok users are open to publicly share their biometric data. There is a greater danger, therefore, for TikTok users to become incorporated in electoral or other databases that can be exploited for varied purposes. Political campaigns and third parties may be eager to collect data on young people as many of them are first-time voters or are still not old enough to vote. As such, they are in the process of creating their political identity and the information they perceive on social media platforms can permeate their eventual ideology. 

TikTok can potentially redefine political communication as a new public arena for civic discourse. While other social media platforms are highly dependent on the friend structure and can thus foster echo-chambers \cite{del2016echo,boutyline2017social}, TikTok's open structure may allow a more cross-partisan dialogue. Whether this assumption holds can only be answered through future research. Even if the hypothesis is proven true, the caveat that political confrontation can be counterproductive exists, and maybe especially applicable to a platform that is highly focused on the virality and humor of its content. Videos with sarcastic and ridiculing content can exacerbate bullying and other harmful behaviors that particularly afflict teenagers \cite{alim2017cyberbullying}. The research community should conduct further analyses that include psychological examinations of the influence exercised by the platform on the youth.

\subsection*{Ethical Concerns}

While conducting this study, we encountered serious ethical questions that must be taken into consideration. First, we crawled TikTok and explicitly collected public data, a portion of which concerned the behavior of young users. Given their age, young users may not be yet be fully aware of the consequences of putting themselves in the public sphere. To maintain data protection, we deleted the collected materials after the analyses. However, we preserved the video ids to allow the replication of this study.  These ids can be found in our GitHub repository\footnote{https://github.com/JuanCarlosCSE/TikTok/blob/master/tiktok\_ids.txt}.

We encountered further ethical issues during the use of the Microsoft Azure image recognition APIs for the detection of age and gender. First, researchers have shown that such algorithms can potentially misclassify minorities and social groups \cite{buolamwini2018gender}. Second, the algorithms for gender classification only provide binary male/female inferences and automatically neglect the existence of other genders. These issues should be kept in mind when this study is read and addressed in the future to promote ethical and inclusive research. 

\section{Conclusion}
In this paper, we studied political communication on TikTok for the first time. We focused on videos related to US politics and evaluated textual, aural and visual information extracted from them. We analyzed the different levels of communication made possible by the platform design and especially concentrated on TikTok's unique duet feature. We then investigated the duet interactions from pro-Republican and pro-Democrat users. In our sample, we find a larger collection of Republican videos, which, on average, attracted more interactions than Democratic videos. Although we find that Democratic users are younger than Republican users, the majority of the users in our data are below 40 years old. We observed that Republican users generated duet videos from users who professed the same ideology more often, whereas Democratic users interacted more with cross-ideological users. Irrespective of their political preferences, however, partisan users expressed themselves in similar ways. Finally, we identified that political content appears to be a relevant aspect of TikTok's ecosystem. Further research is needed to understand how political content is disseminated on this novel social media platform. It would be especially beneficial if prospective studies examined the platform's design and its recommendation system because they are pivotal to the creation of user communities and the shaping of political interactions. Only through rigorous auditing can it be ensured that TikTok represents an open and unbiased arena for political communication.

\bibliographystyle{ACM-Reference-Format}
\bibliography{main}
\end{document}